\documentclass[final,5p,times,twocolumn,sort&compress]{elsarticle}

\usepackage[utf8]{inputenc} 
\usepackage{amsmath}
\usepackage{amssymb}
\usepackage[nolist]{acronym} 
\usepackage[separate-uncertainty=true]{siunitx}

\usepackage{booktabs}   
\usepackage{tabulary}   
\usepackage{ifthen}

\usepackage{microtype}
\usepackage{caption}
\usepackage{subcaption} 
\PassOptionsToPackage{hyphens}{url}\usepackage[hidelinks]{hyperref}

\usepackage{orcidlink}

\usepackage{tikz}
\usetikzlibrary{arrows.meta,calc}

\DeclareSIUnit\year{yr}  
\DeclareSIUnit\ckky{cts/(keV \x kg \x yr)}  

\DeclareSIUnit\angstrom{\text{Å}}
\DeclareSIUnit\bar{bar}
\DeclareSIUnit\barn{b}

\newcommand{\me}{\mathrm{e}}

\graphicspath{ {./figures/} {./} }
\newcommand{\arft}{$^{42}$Ar}
\newcommand{\kft}{$^{42}$K}

\newcommand{\arnat}{$^{\mathrm{nat}}$Ar}
\newcommand{\aar}{atm-Ar}
\newcommand{\uar}{ugr-Ar}

\newcommand{\ovbb}{$0\nu\beta\beta$}

\newcommand{\GERDA}{\textsc{Gerda}}

\newcommand{\legend}{\textsc{Legend}}

\newcommand{\scarf}{\textsc{Scarf}}

\newcommand{\SIasym}[4]{#1 \, ^{+ #2} _{- #3} \, \unit{#4}}

\newboolean{preprintmode}
\setboolean{preprintmode}{false}
\ifthenelse{\boolean{preprintmode}}{}{}

\newboolean{linenrs} 
\ifthenelse{\boolean{preprintmode}}{\setboolean{linenrs}{true}}{\setboolean{linenrs}{false}}

\ifthenelse{\boolean{linenrs}}{\usepackage{lineno}}{\relax}

\journal{Nuclear Instruments and Methods in Physics Research A}

\begin{document}

\begin{frontmatter}



\title{\texorpdfstring{$^{42}$Ar}{Ar-42} Production and Injection to a Liquid Argon Environment for Background Mitigation Studies 
}


\author[TUPH]{Mario Schwarz \orcidlink{0000-0002-8360-666X}\corref{cor1}}
\ead{mario.schwarz@tum.de}
\cortext[cor1]{Corresponding author.}
\author[TUPH]{Christoph Vogl~\orcidlink{0000-0001-9934-5401}}
\author[TUPH]{Niko N. P. N. Lay~\orcidlink{0009-0008-2446-4287}}
\author[TUPH]{Tommaso Comellato~\orcidlink{0000-0003-3780-5139}}
\author[TUPH]{Gunther Korschinek~\orcidlink{0000-0001-9537-1688}}
\author[TUPH]{Moritz Neuberger~\orcidlink{0009-0001-8471-9076}}
\author[TUPH]{Oskar Moras}
\author[TUPH]{Patrick Krause~\orcidlink{0000-0002-9603-7865}\fnref{nowSFU}}
\fntext[nowSFU]{Present address: Department of Physics, Simon Fraser University, Burnaby, Canada}
\author[TUPH]{Stefan Schönert~\orcidlink{0000-0001-5276-2881}}

\affiliation[TUPH]{organization={Department of Physics, TUM School of Natural Sciences, Technische
Universität München},
            city={85748 Garching b. München},
            country={Germany}}

\begin{abstract}
Atmosphere-sourced argon contains traces of $^{42}$Ar, whose $\beta^-$-decaying progeny $^{42}$K represents a significant intrinsic background for rare-event experiments using liquid argon (LAr) as detector or shielding medium. 
Understanding and mitigating this background is crucial for current and future large-scale detectors in neutrino and dark-matter physics. 
To enable controlled studies of $^{42}$K behavior and suppression techniques, $^{42}$Ar was produced by irradiating natural argon with \qty{34}{MeV} $^{7}$Li$^{3+}$ ions at the Maier–Leibnitz–Laboratorium tandem accelerator,
using beam currents of \SI{101+-5}{\nano\ampere} and \SI{140+-5}{\nano\ampere},
yielding $(476 \pm 9)\,\mathrm{Bq}$ within two weeks, corresponding to a production rate of $\sim 1 \times 10^{6}$ atoms\,s$^{-1}$. 
The activated argon was injected into the one-ton \textsc{Scarf} cryostat, where two HPGe detectors monitored the subsequent $^{42}$K activity build-up. 
A time-dependent model describing $^{42}$Ar mixing and $^{42}$K equilibration in LAr yielded characteristic mixing time constants between one and two days. 
The established production and injection capability provides a reproducible platform for high-statistics $^{42}$K background studies, essential for developing and validating suppression strategies for next-generation LAr-based rare-event experiments such as \textsc{Legend-1000}.
\end{abstract}



\begin{keyword}
\arft{} \sep \kft{} \sep background suppression



\end{keyword}

\end{frontmatter}

\ifthenelse{\boolean{linenrs}}{\begin{linenumbers}}{\relax}

\section{Introduction}
\label{sec:intro}

The radioactive isotope \arft{} is present in atmosphere-sourced argon (\aar ) and is primarily produced via cosmic activation through the reaction 
$
^{40}\mathrm{Ar}(\alpha,2p)^{42}\mathrm{Ar}
$
initiated by high-energy cosmic-ray particles in the upper atmosphere. 
\arft{} ($Q_\beta = \SI{599}{\kilo\electronvolt}$, $T_{1/2} = \SI{32.9}{\year}$) undergoes $\beta^-$ decay to its fast decaying progeny \kft{} ($Q_\beta = \SI{3525.3}{\kilo\electronvolt}$, $T_{1/2} = \SI{12.4}{\hour}$) as illustrated in Fig.~\ref{fig:Ar42K42decay}.
The latter's decay proceeds predominantly (\qty{82}{\percent}) without $\gamma$ emission, while in \qty{18}{\percent} of cases a prompt \SI{1524.6}{\kilo\electronvolt} $\gamma$ is emitted. 

Given the high $Q_\beta$ value of $^{42}$K, argon sourced from the atmosphere is a background of concern for low-background rare-event searches and large-scale experiments employing argon as detector material. 
This issue is particularly relevant for experiments such as \legend{} \cite{25tk-nctn,LEGEND1000pcdr}, DEAP-3600 \cite{AMAUDRUZ20191}, DarkSide-20k \cite{DarkSide-20k:2017zyg}, and DUNE \cite{DUNE:2020lwj}, where argon purity and isotope composition critically affect background modeling and detector design.

\begin{figure}[htb]
    \centering
  \ifthenelse{\boolean{preprintmode}}{
  \includegraphics[width=0.4\linewidth, angle=-90]{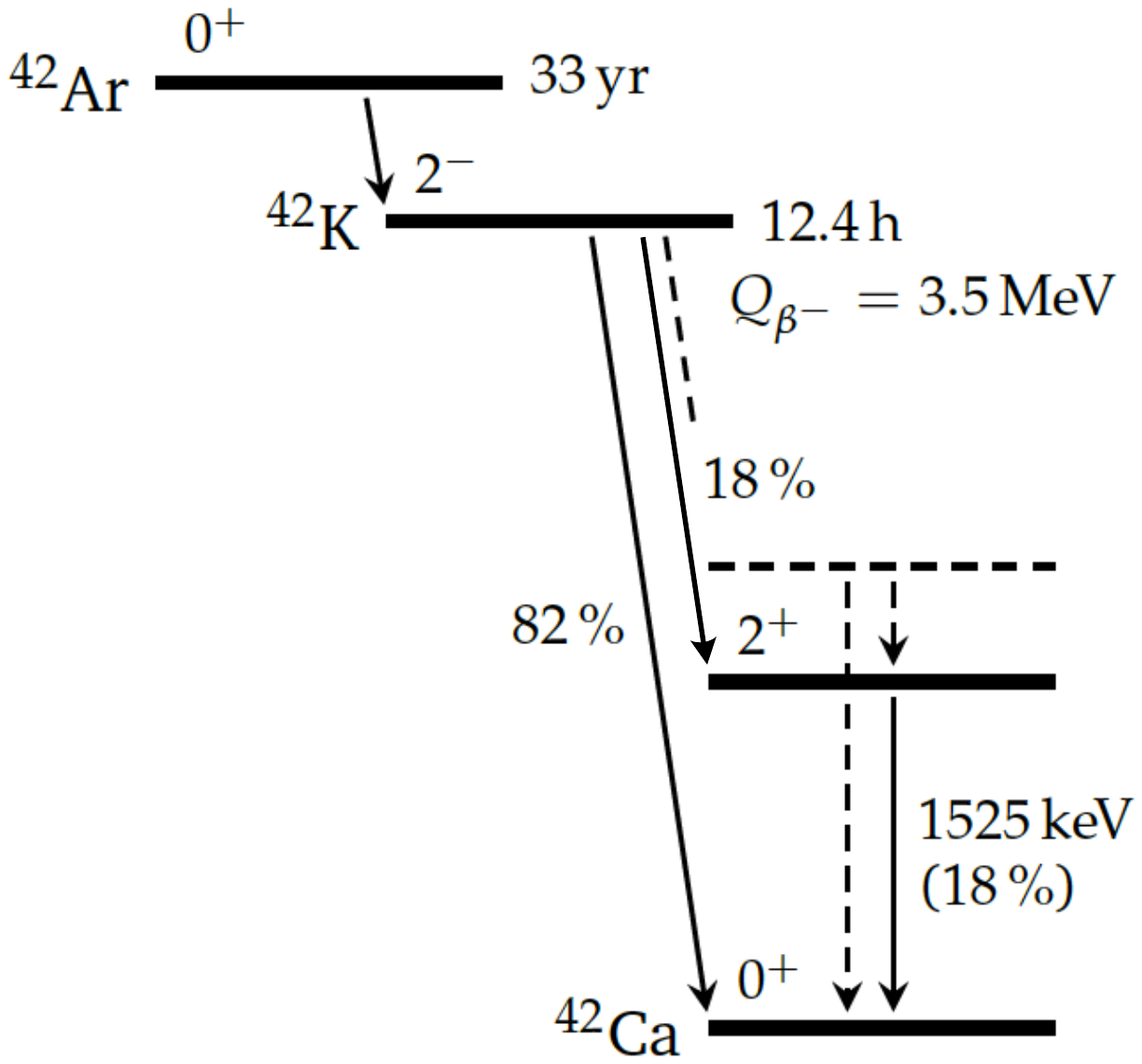}
  }{
  \includegraphics[width=0.65\linewidth, angle=-90]{Ar42K42decay.pdf}
  }
\caption{Simplified $^{42}$Ar/$^{42}$K decay scheme. Cosmogenic $^{42}$Ar $\beta^-$ decays to $^{42}$K. 
$^{42}$K $\beta^-$ decays to the $^{42}$Ca ground state in \qty{82}{\percent} of the cases, and to the first excited state in \qty{18}{\percent}, which is followed by a \SI{1525}{\keV} $\gamma$ emission.}
\label{fig:Ar42K42decay}     
\end{figure}

The specific activity of \arft{} in \aar\ has been the subject of theoretical and experimental studies. 
Early estimates predicted values of around \SI{100}{\micro \becquerel \per \kg}~\cite{PEURRUNG1997425}, in apparent tension with the first experimental upper limit of $<\qty{40}{\micro \becquerel \per \kg}$ (\qty{90}{\percent} C.L.)~\cite{Ashitkov:2003df}.
Measurements in \GERDA{} found specific activities of 
\qty{71}-\qty{101}{\micro \becquerel \per \kg}~\cite{GSTR-16-004}, prompting a re-analysis of the data of \cite{Ashitkov:2003df} that yielded $\SIasym{68}{17}{32}{\micro \becquerel \per \kg}$~\cite{Barabash:2016uzl}.
DEAP-3600 reported $\qty{40.4 \pm 5.9}{\micro \becquerel \per \kg}$~\cite{DEAP:2019pbk}.

Backgrounds from $^{42}$Ar/$^{42}$K are of particular concern for the \legend{} experiment, which searches for neutrinoless double-beta (\ovbb{}) decay of $^{76}$Ge with high-purity p-type germanium (HPGe) detectors enriched in this isotope and operated in an instrumented liquid argon (LAr) environment providing both cooling and active shielding. 
Since the $\beta$ spectrum of \kft{} extends beyond that of $^{76}$Ge double beta ($\beta\beta$) decay ($Q_{\beta\beta} = \SI{2039}{\kilo\electronvolt}$), \kft{} decays can mimic genuine \ovbb{} signals when $\beta$ particles penetrate the thin n$^+$ contact of HPGe detectors. 

At its next stage, \legend-1000~\cite{LEGEND1000pcdr}, the collaboration aims to further reduce the background level, making \kft{} a critical background source. 
Un\-der\-ground-sourced argon (\uar ), which has been demonstrated to contain substantially reduced levels of $^{39}$Ar~\cite{darksidecollaborationResultsFirstUse2016} and is also expected to be depleted in \arft{}, has therefore been adopted as the baseline option for \legend-1000. 
However, given the technical challenges of \uar\ extraction and its uncertain large-scale availability, alternative suppression strategies must be developed and validated to mitigate this risk.

The significance of the \arft{}/\kft{} background was already observed during the commissioning of \GERDA{} \cite{GERDA:2012qwd}, and subsequently in the \textsc{Large} test facility \cite{LArGe2015}. Unexpectedly high \kft{} intensities were measured, 
partially attributed to drift of ionized \kft{} towards the detectors in the electric fields~\cite{GERDAbkg2013}. 
Mitigation strategies in \GERDA{} included copper mini-shrouds, later replaced by TPB-coated nylon, combined with pulse-shape discrimination and LAr scintillation anti-coincidence~\cite{Lubashevskiy2017}. 
To study these effects, dedicated production of \arft{} was performed at the TUM Maier-Leibnitz Laboratory (MLL) already in 2011–2012, enabling injections into the \SI{1.4}{\tonne} of LAr in \textsc{Large} of up to \SI{84(15)}{\becquerel}~\cite{Lubashevskiy2017}, which increased the count rate in the \SI{1524}{keV} line by a factor of around\footnote{Losses in \arft{} activity during injection lower this measured value \cite{Lubashevskiy2017}.} 240.
These spiked-LAr studies provided sufficient statistics to benchmark suppression techniques at the sensitivity level of \GERDA{}.

\legend{}'s stricter background requirements necessitate new background suppression studies, with enhanced statistics enabled by a higher \arft{} activity.
To this end, we produced a new gaseous \arft{} source at the MLL tandem accelerator and injected it into the one-ton \scarf{} cryostat at TUM. 
The objective of this work is to establish a reproducible procedure for the production and controlled injection of $^{42}$Ar, to quantify the resulting activity, and to characterize the ensuing $^{42}$K equilibrium and mixing dynamics in liquid argon. 
This controlled $^{42}$Ar source enables quantitative studies of $^{42}$K behavior and background formation under realistic detector conditions. 
The results provide the experimental foundation for validating suppression and modeling strategies, and for scaling to higher-activity investigations relevant to \textsc{Legend-1000}. 
Section~2 describes the $^{42}$Ar production and activity determination, Section~3 the injection and monitoring of the $^{42}$K build-up, and Section~4 presents the conclusions and outlook.

\section{\texorpdfstring{$^{42}$Ar}{Ar-42} production at MLL}
\label{sec:production}


\subsection{\texorpdfstring{$^{42}$Ar}{Ar-42} production}

We produced \arft{} at the tandem accelerator of the Maier-Leibnitz-Laboratorium (MLL) in Garching during two beam times: in 2018 and 2019.
Each time, gaseous \arnat{} was irradiated with $^7$Li$^{3+}$ of \SI{34}{\mega\electronvolt} energy. 
The procedure and the setup used are built on the preceding unpublished work by \cite{GSTR12011}. 

Figure~\ref{fig:cell_pic} shows a picture of the cell, which has a length of \SI{40}{\centi\meter}.
A piping and instrumentation diagram (P\&ID) of the cell and the surrounding parts of the setup is provided in Fig.~\ref{fig:cell_pid}.
A \SI{6}{\micro\meter} thin Ti window separates the inner volume of the cell from the vacuum of the beam line.

Initially, all parts of the setup are evacuated, including the cell, the volume between the gate valve and the Ti window, and the stainless steel bottle to which the Ar is transferred after irradiation.
Then, the cell is filled with \arnat{} from a gas cylinder, up to a pressure of \SI{600}{\milli\bar} (absolute).
During the irradiation, the vacuum pump continues to evacuate the transfer bottle via the bypass piping.
The irradiation typically takes around \SI{24}{\hour}, after which the Ar from the cell is condensed into the transfer bottle by filling the dewar containing the bottle with liquid N$_2$.
The cell is not evacuated between irradiations.
Rather, the Ar content is topped off with new \arnat{} to avoid losing \arft{} retained in the cell after filling the bottle.
After the last irradiation of a campaign, the cell is flushed with \arnat{} several times during condensing into the transfer bottle to minimize the loss of \arft{}.

\begin{figure}[htb]
    \centering
  \includegraphics[width=0.9\linewidth]{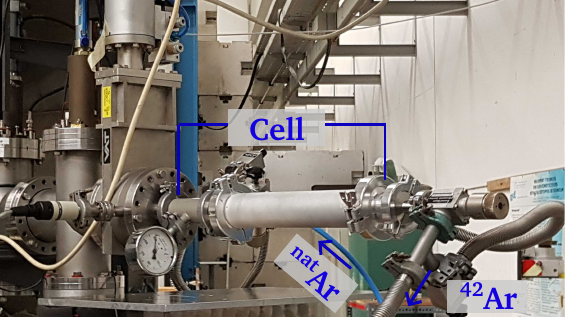}
\caption{
Setup for $^{42}$Ar production at MLL, showing the irradiation cell mounted at the end of the beamline. The $^7$Li$^{3+}$ beam is entering from the left, and irradiating the \qty{40}{\cm} long cell. After irradiation, the activated argon gas is condensed into a LN$_2$-cooled transfer bottle (not shown) below the beamline.}
\label{fig:cell_pic}     
\end{figure}

\begin{figure}[htb]
    \centering
  \includegraphics[width=0.8\linewidth]{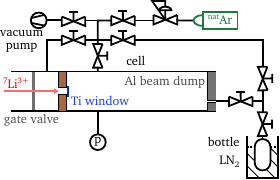}
\caption{P\&ID of the setup used for producing $^{42}$Ar. The gate valve which separates the beam from the irradiation cell is shown alongside the \arnat{} gas cylinder, the \arft{} transfer bottle and the LN$_2$ dewar used to extract the irradiated argon gas from the cell. A vacuum pump is employed to remove air, and several valves define the gas flow.}
\label{fig:cell_pid}     
\end{figure}

During irradiation, the beam current was measured hourly using a Faraday cup.
The mean beam currents are compiled in \autoref{tab:currenttime} together with the elapsed irradiation times.

\begin{table}[tbh]
  \caption{
    The mean current, total elapsed time during irradiations, and number of transport bottles for both beam times. 
    The uncertainty assigned to the mean current derives from the readout accuracy.
  }
  \label{tab:currenttime}
  \begin{center}
  \begin{tabular}{l l l l} \toprule
    beam time   & mean current [\unit{\nano\ampere}] & time [h] & \# bottles \\ \midrule
    2018 & \num{101+-5} & 127.4 & 6 \\
    2019 & \num{140+-5} & 62.3 & 4 \\
  \end{tabular}
  \end{center}
\end{table}

\subsection{Screening of irradiated argon gas}

The transfer bottles were screened individually with an HPGe detector-equipped gamma screening station \cite{HofmannPhD} (see Fig.~\ref{fig:hpge}) at least \SI{11}{days} after the respective irradiation.
Since that corresponds to more than 21 half-lives of $^{42}$K, the contribution from immediately produced $^{42}$K can be neglected, and thus, secular equilibrium can be assumed.
Activities are computed from counts in the \SI{1.5}{\mega\electronvolt} peak from $^{42}$K decay.
A Geant-4 \cite{Geant4} based Monte Carlo simulation \cite{HofmannPhD} implementing the used geometry provides the required detection efficiency.

Measured activities are provided in Tab.~\ref{tab:bottleactive}.
$A_1$ refers to the activity measured shortly after the end of the respective beam times. 
The elapsed times between irradiation and screening range from \SIrange{11}{42}{days}.

\begin{figure}[htb]
    \centering
  \includegraphics[width=0.7\linewidth]{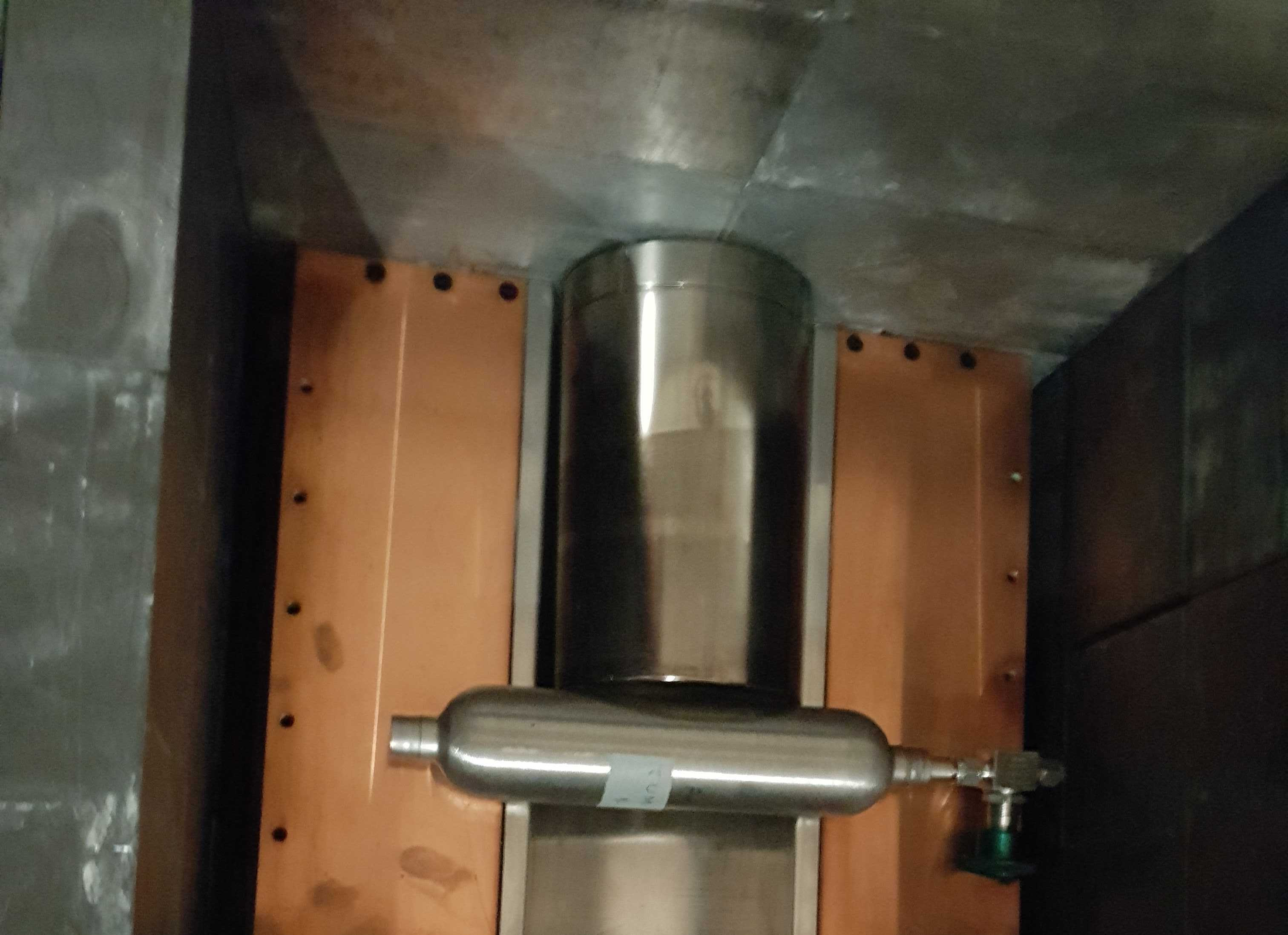}
\caption{A single transfer bottle in the ``GEM'' screening station at TUM. All 10 \arft{} bottles were measured in this station. The first six bottles, filled in 2018, were screened a second time in 2023.}
\label{fig:hpge}     
\end{figure}

The screening was repeated in July 2023 for all bottles filled in the 2018 campaign; the results are reported as $A_2$.
These results can be compared to the expected values $A_\mathrm{calc}$ calculated from $A_1$ using the half-life of \arft{}.
$A_\mathrm{calc}$ refers to the time of the 2nd screening\footnote{For the six bottles filled in 2018, $A_\mathrm{calc}$ refers to the time when the actual 2nd screening happened, while for the four bottles filled in 2019, the reference time point is when the last of the six bottles was re-screened.}.
Since the result for the re-screened bottles agrees well within uncertainties, we conclude that the bottles are tight and that the screening results from the initial screening are not affected by residual $^{42}$K.
Thus, $A_\mathrm{calc}$ is defined here as the reference activity as of July 2023.

\begin{table}[tbh]
  \caption{
    \arft{} activities of the transfer bottles' contents measured with HPGe screening, summed per production campaign.
    $A_1$ was measured some weeks after irradiation, while $A_2$ was measured in July 2023, i.e., around 5 (for the 2018 campaign) or 4 (for the 2019 campaign) years after the respective beam times.
    $A_\mathrm{calc}$ is calculated from $A_1$ using the half-life of \arft{} for July 2023.
  }
  \label{tab:bottleactive}
  \begin{center}
  \begin{tabular}{l l l l} \toprule
    beam time   & $A_1$ [\unit{\becquerel}] & $A_2$ [\unit{\becquerel}] & $A_\mathrm{calc}$ [\unit{\becquerel}] \\ \midrule
    2018 & \num{268+-7} & \num{251+-7} & \num{243+-6} \\
    2019 & \num{208+-6} & --          & \num{192+-5} \\
    \midrule
    $\Sigma$& \num{476+-9} &     --        & \num{434+-8} \\
  \end{tabular}
  \end{center}
\end{table}

\subsection{\texorpdfstring{$^{42}$Ar}{Ar-42} production rate}

The \arft{} production rates $R_\mathrm{meas}$ of $^{42}$Ar nuclei per second are obtained from the measured activities.
Results are derived separately for the 2018 and 2019 campaigns and compiled in \autoref{tab:rates}.

\begin{table}[tbh]
  \caption{
    Rates for the production of \arft{} nuclei from \arnat{} using $^7$Li$^{3+}$ with an initial energy of \SI{34}{\mega\electronvolt}.
    $R_\mathrm{meas}$ is derived from the activity of \arft{} produced in our setup.
    $R_\mathrm{PACE4}$ is the expectation calculated based on the energy-dependent cross section computed with PACE4~\cite{PhysRevC.21.230, PACE4} (see text for details). 
  }
  \label{tab:rates}
  \begin{center}
  \begin{tabular}{l l l} \toprule
    beam time   & $R_\mathrm{meas}$ [\unit{\per\second}] & $R_\mathrm{PACE4}$ [\unit{\per\second}] \\ \midrule
    2018 & \num{8.8+-0.2e5} & \num{3.7+-0.2e5} \\
    2019 & \num{13.9+-0.4e5} & \num{5.1+-0.2e5} \\
  \end{tabular}
  \end{center}
\end{table}

These rates are compared to theoretical predictions based on energy-resolved production cross sections of \arft{} by irradiation of $^{40}$Ar with $^7$Li$^{3+}$ ions.
The cross sections are computed using the fusion evaporation code PACE4~\cite{PhysRevC.21.230, PACE4} included in the LISE++~\cite{TARASOV20084657} framework and presented in \autoref{fig:pace4-cross-sections} (blue data points).

\begin{figure}
    \centering
    \includegraphics[width=\linewidth]{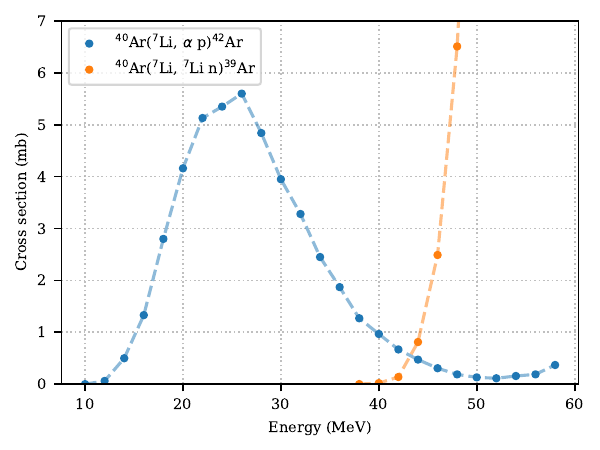}
    \caption{\arft{} and $^{39}$Ar production cross sections for irradiation of $^{40}$Ar with $^7$Li$^{3+}$, predicted by PACE4~\cite{PhysRevC.21.230, PACE4}.}
    \label{fig:pace4-cross-sections}
\end{figure}

The Bethe-Bloch formula \cite{Bethe,Bloch} yields the $^7$Li$^{3+}$ ions' energy of around \SI{32.7}{\mega\electronvolt} immediately after entering the cell and around \SI{0.6}{\mega\electronvolt} at the end of the cell.
Thus, the ions' energy passes the peak cross section, motivating the choice of the initial beam energy.
Using the distance-dependent energy $E(x)$ computed with the Bethe-Bloch formula gives the calculated production rate $R_\mathrm{PACE4}$ of $^{42}$Ar nuclei per second from the PACE4 result via\footnote{The formula is derived from $R=nJl\sigma$, where $n$ is the number density of target nuclei, $J$ is the beam's particle current, $l$ is the length of the cell and $\sigma$ is the production cross section.}
\begin{equation}
    R_\mathrm{PACE4} = \frac{N_A \rho}{M_\mathrm{Ar}} \frac{I}{3 e} \int_0^l{\sigma_\mathrm{PACE4}(E(x)) \mathrm{d}x},
\end{equation}
where $e$ is the elementary charge, $I$ is the mean beam current, $M_\mathrm{Ar}$ is the molecular mass of \arnat{}, $\rho$ is the mass density of the \arnat{} gas within the cell, $N_A$ is Avogadro's constant, $l$ is the length of the cell (\SI{40}{\centi\meter}), and $\sigma_\mathrm{PACE4}$ the energy dependent cross section.
The integral is solved numerically.
The results are listed in \autoref{tab:rates} and hint towards an under-prediction of the production cross section by PACE4.

Furthermore, the theoretical cross section of $^{39}$Ar (orange data points in \autoref{fig:pace4-cross-sections}) is computed, which is an unwanted contaminant as it is a beta-decaying radioisotope that can induce pile-up events in the HPGe detectors.
The chosen beam energy prevents the production of $^{39}$Ar.

\section{Injection of \texorpdfstring{$^{42}$Ar}{Ar-42} into SCARF}
\label{sec:injection}

\subsection{Experimental setup}

The Subterranean Cryogenic Argon Research Facility (\textsc{Scarf}) \cite{wiesingerTUMLiquidArgon2014} is a \qty{1}{t} liquid argon (LAr) cryostat mainly dedicated to R\&D for the \textsc{Gerda} \cite{krauseNewLiquidArgon2019} and \textsc{Legend} \cite{krauseShiningLightBackgrounds2023,schwarzLiquidArgonInstrumentation2021,m.haranczykPurificationLargeVolume2021} experiments, and LAr studies \cite{schwarzTracingImpuritiesIlluminating2024,voglScintillationOpticalProperties2022}. 
A lock system and passive cooling by LN$_2$ enable operation with the same LAr content for over a year while maintaining sufficient purity.
\textsc{Scarf} is located beneath a shallow overburden of a few meters in the underground laboratory (UGL) of the Technical University of Munich (TUM). 

We immersed a string of two p-type HPGe detectors in the LAr of \textsc{Scarf} to obtain data during and around the \arft{} injection.
The detectors are of different geometry; one is a broad energy germanium detector (BEGe), as used in \textsc{Gerda} and \textsc{Legend-200} \cite{agostiniCharacterization30$$^76$$Ge2019}. The other one features the newer inverted coaxial (IC) design \cite{agostiniCharacterizationInvertedCoaxial2021}, which is the 
detector geometry to be used in \textsc{Legend-1000} \cite{LEGEND1000pcdr}.
Signals from the HPGes are amplified with \textsc{Gerda}-like CC3 charge-sensitive amplifiers and the readout is single-ended. 
Waveforms are digitized by a Struck SIS3316 flash analog-to-digital converter.
It is triggered by a signal in either HPGe detector, in which case both detectors are read out simultaneously.
Data is analyzed offline with a pipeline based on C++/ROOT \cite{brunROOTObjectOriented1997}, with the digital signal processing performed by the GELATIO \cite{agostiniGELATIOGeneralFramework2011} software.
The high-level analysis is implemented in Python.
The energy scale is determined by regular $^{228}$Th calibrations, which provide several gamma peaks up to \qty{2.6}{\mega \eV},
and the energy resolution 
has typical values for the full width at half maximum (FWHM) from \qty{3.5}{\keV} to \qty{4}{\keV} 
at the \qty{2615}{\keV} full energy peak.

\subsection{Injection of \texorpdfstring{$^{42}$Ar}{Ar-42} into \textsc{Scarf}}

While the HPGe detectors are biased
and the data acquisition (DAQ) is active, we inject gaseous $^{42}$Ar into the ullage of \textsc{Scarf}.
A simplified piping and instrumentation diagram (P\&ID) of the injection setup and \textsc{Scarf} is provided in \autoref{fig:injection_pid}.
    \begin{figure}[htb]
        \centering
        \includegraphics[width=0.8\linewidth]{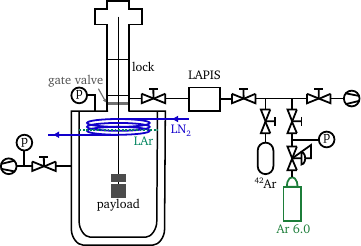}
        \caption{P\&ID of the \arft{} injection setup together with a simplified view of \scarf{} and the payload (2 HPGe detectors) immersed in LAr. A lock system allows for the exchange and maintenance of setups without contaminating the LAr. \arft{} is transferred from the transport bottle into the cryostat through the purification system \textsc{Lapis}. The Ar 6.0 is used to pressurize the transport bottle before the injection, and a turbo-molecular pump evacuates the piping between injections.}
        \label{fig:injection_pid}     
    \end{figure}

The gas lines were evacuated before each injection using a turbo-molecular pump to prevent residual contamination.
Commercial Ar 6.0 is used to pressurize the $^{42}$Ar bottle to around \qty{30}{bar}, to push $^{42}$Ar-rich gas into the gas phase of \textsc{Scarf}. 
Since contamination with air could not be excluded during the $^{42}$Ar production or subsequent storage, the gas from the $^{42}$Ar bottles was directed through the Liquid Argon Purification Instrument for \textsc{Scarf} (\textsc{Lapis})~\cite{voglLiquidphaseLoopmodeArgon2024} before entering the cryostat’s ullage volume.
\textsc{Lapis} removes oxygen, water, and limited amounts of nitrogen from both liquid and gaseous argon streams. 
Because potassium is non-volatile under these conditions, any $^{42}$K already present in the bottles is expected to be trapped within \textsc{Lapis}, preventing its transfer into the cryostat.
Once the $^{42}$Ar enters the ullage volume, it gradually mixes into the liquid argon until a concentration equilibrium is established between the gas and liquid phases. 
Through its $\beta^-$ decay, $^{42}$Ar subsequently produces $^{42}$K, which reaches secular equilibrium within a few days.

\subsection{Energy spectrum}
Before the injection of $^{42}$Ar, a background dataset was recorded to characterize the intrinsic and environmental radioactivity of the setup. 
The corresponding energy spectrum measured with the BEGe detector is shown in purple in \autoref{fig:gamma_spectra}. 
Several common gamma peaks are visible, for example, the \qty{511}{\keV} and \qty{2615}{\keV} lines. 
No indications of unexpected contamination were observed.
    \begin{figure*}[htb]
        \centering
        \includegraphics[width=\linewidth]{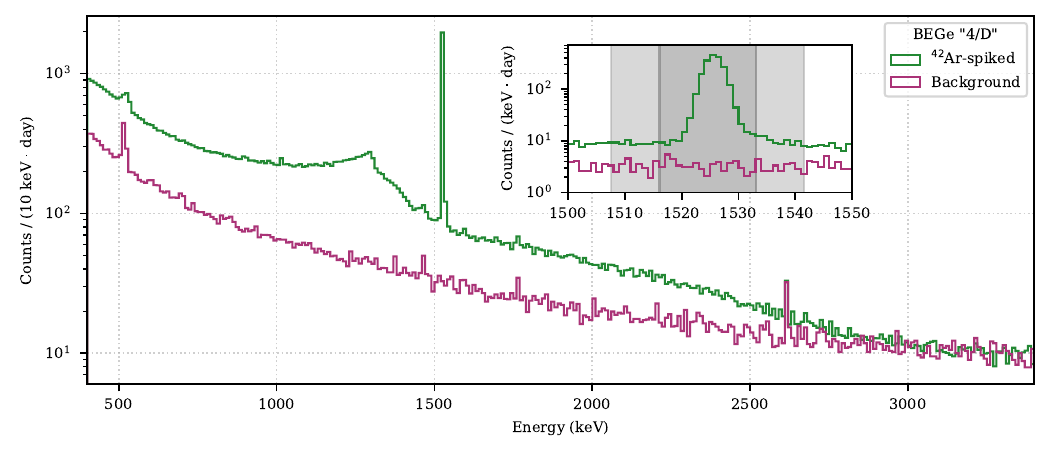}
        \caption{Energy spectrum recorded with a BEGe detector before (purple) and after (green) injection of $^{42}$Ar. The $^{42}$Ar-spiked data were taken after all ten $^{42}$Ar bottles (a total of \qty{434(8)}{\becquerel}) were injected and sufficient time had passed to form a chemical equilibrium between gas and liquid and a secular equilibrium between $^{42}$Ar and $^{42}$K. The spectra are normalized to their respective duration. The $^{42}$Ar-spiked measurement is dominated by the \qty{1525}{\keV} gamma peak of $^{42}$K (see inset). The peak region and the sidebands used in the analysis are shaded in dark and light gray, respectively. Between the gamma line and \qty{3}{\MeV} the $\beta$-decay of $^{42}$K dominates the spectrum.}
        \label{fig:gamma_spectra}     
    \end{figure*}
The green energy spectrum was recorded after the injection of all ten $^{42}$Ar bottles, once both $^{42}$Ar and its daughter $^{42}$K had reached secular equilibrium.
One can see that the \qty{1525}{\keV} line by $^{42}$K is the dominant feature of the spectrum, alongside its pronounced Compton edge. 
While \qty{1525}{\keV} is the main gamma line with $\Gamma = \qty{18.08}{\percent}$ branching ratio, $^{42}$K features additional gamma lines. 
However, their much lower branching ratios and limited statistics render them invisible in this figure. 
In the majority of cases (\qty{81.90}{\percent}), $^{42}$K decays directly into the ground state of $^{42}$Ca.
The electrons from that decay are the origin of the excess events in the $^{42}$Ar-spiked spectrum at energies greater than \qty{1525}{\keV}.
As visible from the figure, the $\beta$ spectrum fuses into the background spectrum at around \qty{3}{\MeV}, even though the $Q_\beta$ value of $^{42}$K is \qty{3.5}{\MeV}. 
This observation has been made before, e.g., in the background decomposition of \textsc{Gerda} \cite{gerdacollaborationModelingGERDAPhase2019} and is predominantly attributed to the energy loss of $\beta$-particles in the HPGe detector's n$^+$ dead layer and transition layer.

\subsection{\texorpdfstring{$^{42}$K}{K-42} build-up}

In addition to the previously studied datasets, the data recorded during the $^{42}$Ar injection and subsequent $^{42}$K build-up were analyzed by monitoring the count rate in the \qty{1525}{\keV} $\gamma$ peak.
The inset in \autoref{fig:gamma_spectra} shows the peak and its surrounding region, with the dark gray area defining the peak window and the adjacent light gray areas representing the sidebands.

The peak region has a width of \qty{17}{\keV} and is centered around the literature position of the peak at \qty{1524.6}{\keV}.
The sidebands extend \SI{8.5}{\keV} in either direction.
The net count rate per day was calculated with \qty{4}{hour} binning, and the result is shown in \autoref{fig:rate_evolution} together with a fit of the dedicated model discussed below. 
Only data from the BEGe detector are presented; the IC detector shows compatible behavior. 
The plot begins at the time of the injection of $^{42}$Ar bottles number 6 and 9, with a summed expected activity of \qty{127.5(5.1)}{\becquerel}.

    \begin{figure}[tb]
        \centering
        \includegraphics[width=\linewidth]{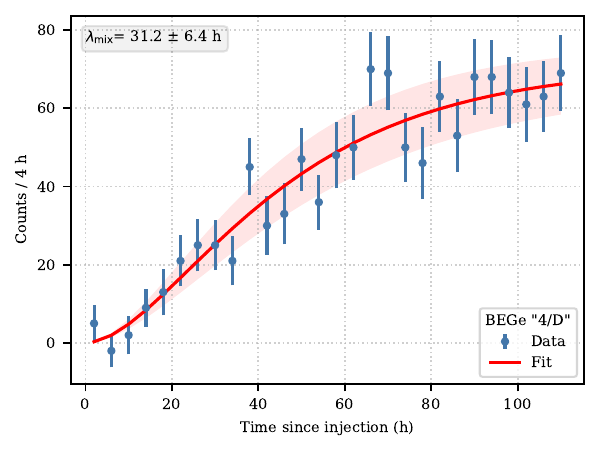}
    \caption{Measured count rate in the \qty{1525}{\keV} gamma line of $^{42}$K after the first injection of $^{42}$Ar into the gas phase of \textsc{Scarf}. Here, bottles 6 and 9 were injected, with a total activity of \qty{127.5(51)}{\becquerel}. The $^{42}$Ar diffuses into the liquid phase and produces $^{42}$K, which is easily detectable via its strong gamma line. A dedicated $^{42}$K build-up model was developed. The result of a fit to data is shown in red; the $1\sigma$ uncertainty band is drawn shaded.}
    \label{fig:rate_evolution}     
    \end{figure}
As expected, the count rate increases quickly after injection and levels off later. 
Once a concentration equilibrium between the gas and liquid phases is established, and a secular equilibrium between $^{42}$Ar and $^{42}$K is formed, the \qty{1525}{\keV} gamma rate is constant.

A model was employed to describe the evolution of the $\gamma$-ray count rate over time, featuring two free parameters: the $^{42}$Ar mixing time constant $\lambda_\mathrm{mix}$ and an efficiency parameter introduced below.
The following overviews the derivation of the model; details are provided in \ref{app:k42-gamma-evolution}.
The model builds on a set of differential equations for $N^{g}_{\mathrm{Ar}}$ and $N^{l}_{\mathrm{Ar}}$, the time-dependent number of $^{42}$Ar atoms the gas and liquid phases, respectively:
\begin{equation}
\begin{split}
\frac{\mathrm{d}N^{g}_{\mathrm{Ar}}}{\mathrm{d}t} &= -(\lambda_\mathrm{mix} + \lambda_\mathrm{Ar}) N^{g}_{\mathrm{Ar}} + \lambda'_\mathrm{mix} N^{l}_{\mathrm{Ar}} \\
\frac{\mathrm{d}N^{l}_{\mathrm{Ar}}}{\mathrm{d}t} &= \lambda_\mathrm{mix} N^{g}_{\mathrm{Ar}} - (\lambda'_\mathrm{mix} + \lambda_\mathrm{Ar}) N^{l}_{\mathrm{Ar}}.
\end{split}
\end{equation}
Both the radioactive decay of \arft{} (decay constant $\lambda_\mathrm{Ar}$) and the transition between the gas and liquid states are included.
In addition to the mixing time constant for the transition from gas to liquid $\lambda_\mathrm{mix}$, the time constant for the reverse process appears ($\lambda'_\mathrm{mix}$).
The boundary conditions are
\begin{equation}
     N^{g}_{\mathrm{Ar}}(0) = A \quad \text{and} \quad N^{l}_{\mathrm{Ar}}(0) = 0,
\end{equation}
with $A$ denoting the initial number of \arft{} atoms, expected to be exclusively in the gas phase.
The differential equations are solved analytically.
Subsequently, the assumptions $\lambda_\mathrm{mix} \gg \lambda_\mathrm{Ar}$, motivated by the  half-life of $^{42}$Ar being \qty{32.9}{yr}, $\lambda_\mathrm{mix} \gg \lambda'_\mathrm{mix}$, derived from the expectation of $N^{l}_{\mathrm{Ar}} \gg N^{g}_{\mathrm{Ar}}$ in equilibrium, and $t \ll 1/\lambda_\mathrm{Ar}$ for the time range of interest allow for simplifications.

A further set of rate equations describes the production of $^{42}$K from the decay of $^{42}$Ar, neglecting the transport of $^{42}$K between the phases, and incorporating the decay into $^{42}$Ca.
In addition to the already mentioned assumptions, the condition $\lambda_\mathrm{K} \gg \lambda_\mathrm{Ar}$ was applied.
Solving this second set of differential equations yields the time-dependent number of $^{42}$K atoms in each phase.
To model the detected $\gamma$ rate, time is binned into windows of width $\Delta t = \qty{4}{h}$, and only the liquid phase is considered to contribute.
This approach yields the final rate of the \qty{1525}{\keV} $\gamma$-line expected from the coupled decay–mix system,
    \begin{equation}
    \label{eq:builduprate}
    \begin{split}
        \text{Rate}(t) &= \xi A \cdot \lambda_\mathrm{Ar} \cdot\Gamma \cdot \lambda_\mathrm{K}\cdot \lambda_\mathrm{mix} \\ 
        &\phantom{={}}\Bigg(
        \frac{1}{\lambda_\mathrm{mix}\lambda_{K}} - \frac{\me^{-\lambda_{K}\Delta t / 2} -\me^{\lambda_{K}\Delta t / 2}}{\lambda_{K}^2(\lambda_{K} - \lambda_\mathrm{mix})\Delta t}\me^{-\lambda_{K}t} \\
        &\phantom{={}}+ \frac{\me^{-\lambda_\mathrm{mix}\Delta t / 2} -\me^{\lambda_\mathrm{mix}\Delta t / 2}}{\lambda_\mathrm{mix}^2(\lambda_{K} - \lambda_\mathrm{mix})\Delta t}\me^{-\lambda_\mathrm{mix}t} \Bigg) \text{,}
    \end{split}
    \end{equation}
where $\xi$ is the detection efficiency and $\Gamma = \qty{18.08}{\percent}$ the relevant branching ratio.
This function has only two free parameters: the $^{42}$Ar mixing time $\lambda_\mathrm{mix}$, and the product $\xi A$,
an effective normalization constant accounting for detection efficiency and total activity.
The model was fit to the data in \autoref{fig:rate_evolution} with good agreement. 
The $1\sigma$ uncertainty band is shown shaded in red. 
The mixing time was determined to be $\lambda_\mathrm{mix} = \qty{32.2(6.4)}{\hour}$. 
Applying the same analysis to the IC detector resulted in a slightly longer mixing time of \qty{40.6(6.8)}{\hour}, which is, however, compatible within uncertainties.

\section{Conclusion and outlook}
\label{sec:concout}

In this paper, we report on the production of \arft{} and its injection into the \textsc{Scarf} cryostat.
We produced \arft{}-rich argon gas at the MLL tandem accelerator by irradiating \arnat{} with a \SI{34}{\mega\electronvolt} $^7$Li$^{3+}$ beam.
The beam energy maximizes the \arft{} production yield while preventing $^{39}$Ar production. 
The obtained \arft{} production rates are around \SI{1e6}{atoms\, \second^{-1}}, surpassing the theoretical calculations based on PACE4 by a factor of more than two.
HPGe screening on all filled transport bottles yields a total \arft{} activity of \SI{476+-9}{\becquerel}, from which \SI{434+-8}{\becquerel} remain at the time of injection.
Diluted in around \SI{1}{\tonne} LAr, a specific activity of 0.43 Bq/kg was achieved, a factor of around 5000 higher than natural atmosphere-sourced LAr\footnote{We consider the results \SI{71}{\becquerel\per\tonne} and \SI{101}{\becquerel\per\tonne} \cite{GSTR-16-004} from \textsc{Gerda} here.}.
In comparison, a factor of 240 was achieved in LArGe \cite{Lubashevskiy2017}.

The build-up of \kft{} activity during and immediately after the injection was measured using a string of two HPGe detectors.
The increase in the \SI{1525}{\kilo\electronvolt} gamma line count rate was successfully modeled, determining the mixing time constant.
Values around \SIrange{32}{41}{\hour} are observed, indicating that stable conditions can be reached days after injection.

A rich R\&D campaign has been carried out in the $^{42}$Ar-spiked LAr of \textsc{Scarf}, focusing on $^{42}$K suppression techniques for \textsc{Legend-1000}. 
The results of these studies will be reported elsewhere. 
The established production and injection capability provides a reproducible platform for high-statistics $^{42}$K background studies, essential for developing and validating suppression concepts toward \textsc{Legend-1000}.

To experimentally demonstrate $^{42}$K suppression factors on the order of \mbox{\num{e5}-\num{e6}}, as required for operating \textsc{Legend-1000} in atmospheric argon, an even higher $^{42}$Ar activity must be produced. 
This can be achieved by irradiating additional batches, either by extending the beam time or by increasing the beam current. 
While the latter approach is generally preferred, it is technically challenging for a tandem accelerator due to the production yield of negative $^7$Li ion beams. 
Therefore, the use of an alternative accelerator type, ideally equipped with an electron-cyclotron resonance (ECR) ion source, is under consideration.

\appendix
\section{Details of \texorpdfstring{$^{42}$K}{K-42} build-up modeling}
\label{app:k42-gamma-evolution}

The exchange of \arft{} atoms between the gas and the liquid phase, together with their radioactive decay, are described using two coupled differential equations:
\begin{equation}
\label{eq:N_gas_dgl}
\begin{split}
\frac{\mathrm{d}N^{g}_{\mathrm{Ar}}}{\mathrm{d}t} &= -(\lambda_\mathrm{mix} + \lambda_\mathrm{Ar}) N^{g}_{\mathrm{Ar}} + \lambda'_\mathrm{mix} N^{l}_{\mathrm{Ar}} \\
\frac{\mathrm{d}N^{l}_{\mathrm{Ar}}}{\mathrm{d}t} &= \lambda_\mathrm{mix} N^{g}_{\mathrm{Ar}} - (\lambda'_\mathrm{mix} + \lambda_\mathrm{Ar}) N^{l}_{\mathrm{Ar}}.
\end{split}
\end{equation}
Here, $N^g_{\mathrm{Ar}}$ and $N^l_{\mathrm{Ar}}$ denote the number of \arft{} atoms in the gas and liquid state, respectively. 
The transfer rate from gas to liquid is referred to as $\lambda_\mathrm{mix}$, while the rate of the reverse transfer is $\lambda'_\mathrm{mix}$; the decay constant is $\lambda_\mathrm{Ar}$.
Injecting into the gas phase leads to the following initial conditions:
\begin{equation}
     N^{g}_{\mathrm{Ar}}(0) = A \quad \text{and} \quad N^{l}_{\mathrm{Ar}}(0) = 0,
\end{equation}
with $A$ being the initial number of \arft{} atoms.

The following solutions are obtained:
\begin{equation}
\begin{split}
    N^{g}_{\mathrm{Ar}}(t) &= \frac{A}{\lambda_\mathrm{mix} + \lambda'_\mathrm{mix}} \left[ \lambda'_\mathrm{mix} \me^{-\lambda_\mathrm{Ar} t} + \lambda_\mathrm{mix} \me^{-(\lambda_\mathrm{mix} + \lambda'_\mathrm{mix} + \lambda_\mathrm{Ar}) t} \right] \\
N^{l}_{\mathrm{Ar}}(t) &= \frac{A \lambda_\mathrm{mix}}{\lambda_\mathrm{mix} + \lambda'_\mathrm{mix}} \left[ \me^{-\lambda_\mathrm{Ar} t} - \me^{-(\lambda_\mathrm{mix} + \lambda'_\mathrm{mix} + \lambda_\mathrm{Ar}) t} \right].
\end{split}
\end{equation}
These can be simplified using $\lambda_\mathrm{mix} \gg \lambda'_\mathrm{mix}$, which derives from \autoref{eq:N_gas_dgl} and the assumption that the vast majority of \arft{} will be in the liquid after equilibrium is reached.
Further, we expect the mixing to proceed much faster than \arft{} decay due to its long half-life, i.e. $\lambda_\mathrm{mix} \gg \lambda_{Ar}$.
The time range of interest is $t \ll 1/\lambda_\mathrm{Ar}$.
The simplified solutions read:
\begin{equation}
\label{eq:Ar_gas_solv_simpl}
\begin{split}
N^{g}_{\mathrm{Ar}}(t) &= A \me^{-\lambda_\mathrm{mix}t} \\
N^{l}_{\mathrm{Ar}}(t) &= A \left[ 1 - \me^{-\lambda_\mathrm{mix} t} \right].
\end{split}
\end{equation}

As \arft{} transfers and mixes in the \textsc{Scarf} cyostat, $^{42}$K builds up in the gas and liquid phases due to $^{42}$Ar decay. 
Neglecting the contribution due to a direct transfer of $^{42}$K between the phases, the rate of change of the number of excess $^{42}$K in the gas and liquid phase follows
\begin{equation}
    \label{eq:k-42_rate_eq}
    \begin{split}
        \frac{\mathrm{d}N^{g}_{\mathrm{K}}}{\mathrm{d}t} &= \lambda_{\mathrm{Ar}}N^{g}_{\mathrm{Ar}} - \lambda_{\mathrm{K}}N^{g}_{\mathrm{K}}\\
        \frac{\mathrm{d}N^{l}_{\mathrm{K}}}{\mathrm{d}t} &= \lambda_{\mathrm{Ar}}N^{l}_{\mathrm{Ar}} - \lambda_{\mathrm{K}}N^{l}_{\mathrm{K}},
    \end{split}
\end{equation}
where $\lambda_{\text{K}}$ is the decay constant of $^{42}$K. The first terms account for the build-up due to the decay of \arft{} into $^{42}$K, and the second terms handle the decrease due to $^{42}$K decay into $^{42}$Ca.

Substituting \autoref{eq:Ar_gas_solv_simpl} into \autoref{eq:k-42_rate_eq} and solving the differential equations with the boundary conditions ${N^{g}_{\mathrm{K}}(t=0) = N^{l}_{\mathrm{K}}(t=0) = 0}$ yield
\begin{equation}
    \label{eq:n_K_vs_t}
    \begin{split}
        N^{g}_{\mathrm{K}}(t) =& \lambda_{\mathrm{Ar}}A \frac{\me^{-\lambda_{\mathrm{mix}}t} - \me^{-\lambda_{\mathrm{K}}t}}{\lambda_{\mathrm{K}} - \lambda_{\mathrm{mix}}}\\
        N^{l}_{\mathrm{K}}(t) =& \lambda_{\mathrm{mix}}\lambda_{\mathrm{Ar}}A\Bigg(\frac{1}{\lambda_{\mathrm{mix}}\lambda_{\mathrm{K}}} + \frac{1}{\lambda_{\mathrm{K}}\left(\lambda_{\mathrm{K}}-\lambda_{\mathrm{mix}}\right)}\me^{-\lambda_{\mathrm{K}}t}\\
        &- \frac{1}{\lambda_{\mathrm{mix}}\left(\lambda_{\mathrm{K}} - \lambda_{\mathrm{mix}}\right)}\me^{-\lambda_{\mathrm{mix}}t}\Bigg)\mathrm{.}
    \end{split}
\end{equation}

Finally, the decay of $^{42}$K into the stable $^{42}$Ca isotope results in $^{42}$Ca build-up in the liquid phase. The number of excess $^{42}$Ca in the liquid phase in coincidence with a \SI{1.5}{MeV} gamma emission is given by
\begin{equation}
    \label{eq:Ca-42_vs_t}
    N^{l}_{\mathrm{Ca, }\gamma}(t) = \Gamma\cdot \lambda_{\mathrm{K}}\int_0^t N^{l}_{\mathrm{K, }\gamma}(\tilde{t})\ \mathrm{d}\tilde{t},
\end{equation}
with $\Gamma$ the branching ratio of the \SI{1.5}{MeV} gamma emission. The time evolution of the count rate of the 1.5 MeV line is calculated in equally spaced time bins, with a width of $\Delta t = t_2 - t_1$, starting from the time \arft{} was injected into the \textsc{Scarf} cryostat. Here, $t_1$ and $t_2$ are the corresponding bin edges. Each count rate value is then assigned a timestamp $t = \frac{t_1 + t_2}{2}$. Assuming that only gammas from $^{42}$K decays occurring in the liquid phase are detected, the measured count rate in each time bin is given by
\begin{equation}
    \label{eq:count_rate_app}
    \begin{split}
        \mathrm{Rate}(t) =& \xi \frac{N^{l}_{\mathrm{Ca, }\gamma}(t_2) - N^{l}_{\mathrm{Ca, }\gamma}(t_1)}{\Delta t}\\
        =& \xi A \cdot \lambda_\mathrm{Ar} \cdot\Gamma \cdot \lambda_\mathrm{K}\cdot \lambda_\mathrm{mix} \\ 
        &\phantom{={}}\Bigg(
        \frac{1}{\lambda_\mathrm{mix}\lambda_{K}} - \frac{\me^{-\lambda_{K}\Delta t / 2} -\me^{\lambda_{K}\Delta t / 2}}{\lambda_{K}^2(\lambda_{K} - \lambda_\mathrm{mix})\Delta t}\me^{-\lambda_{K}t} \\
        &\phantom{={}}+ \frac{\me^{-\lambda_\mathrm{mix}\Delta t / 2} -\me^{\lambda_\mathrm{mix}\Delta t / 2}}{\lambda_\mathrm{mix}^2(\lambda_{K} - \lambda_\mathrm{mix})\Delta t}\me^{-\lambda_\mathrm{mix}t} \Bigg)\mathrm{,}
    \end{split}
\end{equation}
where $\xi$ is the detection efficiency.

\section*{Acknowledgements}
We thank the beam operators at the Maier-Leibnitz-Laboratorium for their support.
We further acknowledge the work of Du\v{s}an Budjá\v{s} and other persons involved in the 2011 and 2012 campaigns.

This work was supported by the Deutsche Forschungsgemeinschaft (DFG, Excellence Cluster \textsc{Origins} EXC-2094–390783311 and Collaborative Research Center SFB 1258 “Neutrinos and Dark Matter in Astro- and Particle Physics”) and by the Maier-Leibnitz-Laboratorium Garching.

\ifthenelse{\boolean{linenrs}}{\end{linenumbers}}{\relax}
\bibliographystyle{elsarticle-num} 
\bibliography{main.bib}

@article{25tk-nctn,
    author = "{LEGEND Collaboration} and Acharya, H. and others",
    title = "{First Results on the Search for Lepton Number Violating Neutrinoless Double Beta Decay with the LEGEND-200 Experiment}",
    eprint = "2505.10440",
    archivePrefix = "arXiv",
    primaryClass = "hep-ex",
    doi = "10.1103/25tk-nctn",
    month = "5",
    year = "2025"
}

@article{LEGEND1000pcdr,
    author = "{LEGEND Collaboration} and Abgrall, N. and others",
    collaboration = "LEGEND",
    title = "{The Large Enriched Germanium Experiment for Neutrinoless $\beta\beta$ Decay}: {LEGEND-1000 Preconceptual Design Report}",
    eprint = "2107.11462",
    archivePrefix = "arXiv",
    primaryClass = "physics.ins-det",
    month = "7",
    year = "2021",
}

@article{AMAUDRUZ20191,
    title = {Design and construction of the {DEAP-3600} dark matter detector},
    journal = {Astroparticle Physics},
    volume = {108},
    pages = {1-23},
    year = {2019},
    issn = {0927-6505},
    doi = {10.1016/j.astropartphys.2018.09.006},
    author = {{DEAP Collaboration} and Amaudruz, P.-A. and others},
}

@article{DarkSide-20k:2017zyg,
    author = "{DarkSide-20k Collaboration} and Aalseth, C. E. and others",
    collaboration = "DarkSide-20k",
    title = "{DarkSide-20k: A 20 tonne two-phase LAr TPC for direct dark matter detection at LNGS}",
    eprint = "1707.08145",
    archivePrefix = "arXiv",
    primaryClass = "physics.ins-det",
    reportNumber = "FERMILAB-PUB-17-298-PPD",
    doi = "10.1140/epjp/i2018-11973-4",
    journal = "Eur. Phys. J. Plus",
    volume = "133",
    pages = "131",
    year = "2018",
}

@article{DUNE:2020lwj,
    author = "{DUNE Collaboration} and Abi, Babak and others",
    collaboration = "DUNE",
    title = "{Deep Underground Neutrino Experiment (DUNE), Far Detector Technical Design Report, Volume I Introduction to DUNE}",
    eprint = "2002.02967",
    archivePrefix = "arXiv",
    primaryClass = "physics.ins-det",
    reportNumber = "FERMILAB-PUB-20-024-ND, FERMILAB-DESIGN-2020-01",
    doi = "10.1088/1748-0221/15/08/T08008",
    journal = "JINST",
    volume = "15",
    number = "08",
    pages = "T08008",
    year = "2020",
}

@article{PEURRUNG1997425,
    title = {Expected atmospheric concentration of 42Ar},
    journal = {Nucl. Inst. Meth. A},
    volume = {396},
    number = {3},
    pages = {425-426},
    year = {1997},
    issn = {0168-9002},
    doi = {10.1016/S0168-9002(97)00819-X},
    author = {A.J Peurrung and T.W Bowyer and R.A Craig and P.L Reeder},
}

@ARTICLE{Ashitkov:2003df,
    author = {{Ashitkov}, V.~D. and {Barabash}, A.~S. and {Belogurov}, S.~G. and others},
    title = "{Liquid Argon Ionization Detector for Double Beta Decay Studies}",
    journal = {arXiv e-prints},
    year = 2003,
    month = sep,
    archivePrefix = {arXiv},
    eprint = {nucl-ex/0309001},
    primaryClass = {nucl-ex},
}

@article{Barabash:2016uzl,
    author = "Barabash, A. S. and Saakyan, R. R. and Umatov, V. I.",
    title = "{On concentration of $^{42}$Ar in the Earth's atmosphere}",
    eprint = "1609.08890",
    archivePrefix = "arXiv",
    primaryClass = "nucl-ex",
    doi = "10.1016/j.nima.2016.09.042",
    journal = "Nucl. Instrum. Meth. A",
    volume = "839",
    pages = "39--42",
    year = "2016"
}

@article{Lubashevskiy2017,
    author = "Lubashevskiy, A. and others",
    title = "{Mitigation of $^{42}$Ar/$^{42}$K background for the GERDA Phase II experiment}",
    eprint = "1708.00226",
    archivePrefix = "arXiv",
    primaryClass = "physics.ins-det",
    doi = "10.1140/epjc/s10052-017-5499-9",
    journal = "Eur. Phys. J. C",
    volume = "78",
    number = "1",
    pages = "15",
    year = "2018",
}

@article{GERDA:2012qwd,
    author = "{GERDA Collaboration} and Ackermann, K. H. and others",
    collaboration = "GERDA",
    title = "{The GERDA experiment for the search of $0\nu\beta\beta$  decay in $^{76}$Ge}",
    eprint = "1212.4067",
    archivePrefix = "arXiv",
    primaryClass = "physics.ins-det",
    doi = "10.1140/epjc/s10052-013-2330-0",
    journal = "Eur. Phys. J. C",
    volume = "73",
    number = "3",
    pages = "2330",
    year = "2013",
}

@article{LArGe2015,
    author = {Agostini, Matteo and Barnabé-Heider, M. and Budjas, Dusan and others},
    year = {2015},
    month = {10},
    pages = {},
    title = {{{LArGe}}: active background suppression using argon scintillation for the {{GERDA}} $0\nu\beta\beta$-experiment},
    volume = {75:506},
    journal = {European Physical Journal C},
    doi = {10.1140/epjc/s10052-015-3681-5},
}

@article{GERDAbkg2013,
    author = "{GERDA Collaboration} and Agostini, M. and others",
    collaboration = "GERDA",
    title = "{The background in the $0 \nu \beta \beta$ experiment GERDA}",
    eprint = "1306.5084",
    archivePrefix = "arXiv",
    primaryClass = "physics.ins-det",
    doi = "10.1140/epjc/s10052-014-2764-z",
    journal = "Eur. Phys. J. C",
    volume = "74",
    number = "4",
    pages = "2764",
    year = "2014",
}

@misc{GSTR12011,
    author = "Budjas, D. and Janicskó-Csáthy, J. and Korschinek, G. and others",
    title = "{Production of $^{42}$Ar for high intensity $^{42}$K background studies with \textsc{LArGe}}",
    note = {{GERDA} Collaboration, internal report},
    year = "2012",
}

@article{DEAP:2019pbk,
    author = "{DEAP Collaboration} and Ajaj, R. and others",
    collaboration = "DEAP",
    title = "{Electromagnetic backgrounds and potassium-42 activity in the DEAP-3600 dark matter detector}",
    eprint = "1905.05811",
    archivePrefix = "arXiv",
    primaryClass = "nucl-ex",
    doi = "10.1103/PhysRevD.100.072009",
    journal = "Phys. Rev. D",
    volume = "100",
    number = "7",
    pages = "072009",
    year = "2019",
}

@article{Geant4,
    title = {Geant4—a simulation toolkit},
    journal = {Nucl. Inst. Meth. A},
    volume = {506},
    number = {3},
    pages = {250-303},
    year = {2003},
    issn = {0168-9002},
    doi = {10.1016/S0168-9002(03)01368-8},
    author = {S. Agostinelli and J. Allison and K. Amako and others},
}

@phdthesis{HofmannPhD,
    author = {Hofmann, Martin Alexander},
    title = {Liquid Scintillators and Liquefied Rare Gases for Particle Detectors},
    year = {2012},
    school = {Technische Universität München},
    pages = {294},
    language = {en},
    url = {https://mediatum.ub.tum.de/1115726},
}

@article{agostiniCharacterization30$$^76$$Ge2019,
    title = {Characterization of 30 $^{76}${{Ge}} Enriched {{Broad Energy Ge}} Detectors for {{GERDA Phase II}}},
    author = {{GERDA Collaboration} and Agostini, M. and others},
    year = {2019},
    month = nov,
    journal = {Eur. Phys. J. C},
    volume = {79},
    number = {11},
    pages = {978},
    issn = {1434-6052},
    doi = {10.1140/epjc/s10052-019-7353-8},
}

@article{agostiniCharacterizationInvertedCoaxial2021,
    title = {Characterization of Inverted Coaxial $^{76}${{Ge}} Detectors in {{GERDA}} for Future Double-$\beta\beta$ decay Experiments},
    author = {{GERDA Collaboration} and Agostini and others},
    year = {2021},
    month = jun,
    journal = {Eur. Phys. J. C},
    volume = {81},
    number = {6},
    pages = {505},
    issn = {1434-6052},
    doi = {10.1140/epjc/s10052-021-09184-8},
    langid = {english},
}

@article{agostiniGELATIOGeneralFramework2011,
    title = {{{GELATIO}}: A General Framework for Modular Digital Analysis of High-Purity {{Ge}} Detector Signals},
    shorttitle = {{{GELATIO}}},
    author = {Agostini, M and Pandola, L and Zavarise, P and Volynets, O},
    year = {2011},
    month = aug,
    journal = {J. Inst.},
    volume = {6},
    number = {08},
    pages = {P08013-P08013},
    issn = {1748-0221},
    doi = {10.1088/1748-0221/6/08/P08013},
    urldate = {2019-09-09},
}

@article{brunROOTObjectOriented1997,
    title = {{{ROOT}} --- {{An}} Object Oriented Data Analysis Framework},
    author = {Brun, Rene and Rademakers, Fons},
    year = {1997},
    month = apr,
    journal = {Nucl. Inst. Meth. A},
    volume = {389},
    number = {1},
    pages = {81--86},
    issn = {0168-9002},
    doi = {10.1016/S0168-9002(97)00048-X},
}

@article{gerdacollaborationModelingGERDAPhase2019,
    author = "{GERDA Collaboration} and Agostini, M. and others",
    title = "{Modeling of GERDA Phase II data}",
    eprint = "1909.02522",
    archivePrefix = "arXiv",
    primaryClass = "nucl-ex",
    doi = "10.1007/JHEP03(2020)139",
    journal = "JHEP",
    volume = "03",
    pages = "139",
    year = "2020"
}

@phdthesis{krauseNewLiquidArgon2019,
    type = {M.{{Sc}}. Thesis},
    title = {The {{New Liquid Argon Veto}} of {{GERDA}}},
    author = {Krause, Patrick},
    year = {2019},
    address = {Garching},
    langid = {english},
    school = {Techniche Universit{\"a}t M{\"u}nchen},
}

@phdthesis{krauseShiningLightBackgrounds2023,
    title = {Shining {{Light}} on {{Backgrounds}}},
    author = {Krause, Patrick},
    year = {2023},
    address = {Garching},
    langid = {english},
    school = {Techniche Universit{\"a}t M{\"u}nchen},
}

@article{m.haranczykPurificationLargeVolume2021,
    author = "Haranczyk, Malgorzata and others",
    title = "{Purification of large volume of liquid argon for LEGEND-200}",
    doi = "10.22323/1.380.0102",
    journal = "PoS",
    volume = "PANIC2021",
    pages = "102",
    year = "2022",
}

@inproceedings{schwarzLiquidArgonInstrumentation2021,
    ids = {marioschwarzLiquidArgonInstrumentation2021},
    title = {Liquid {{Argon Instrumentation}} and {{Monitoring}} in {{LEGEND-200}}},
    booktitle = {{{ANIMMA}} 2021 {{Conference Proceedings}}},
    author = {Schwarz, Mario and Krause, Patrick and Fomina, Maria and others},
    year = {2021},
    month = aug,
    publisher = {EPJ Web of Conferences},
    doi = {10.1051/epjconf/202125311014},
}

@phdthesis{schwarzTracingImpuritiesIlluminating2024,
    title = {Tracing Impurities and Illuminating Their Impact: {{Surveying}} and Characterizing Liquid Argon with {{LLAMA}} for {{LEGEND}} and Beyond},
    author = {Schwarz, Mario},
    year = {2024},
    address = {Garching},
    langid = {english},
    school = {Techniche Universit{\"a}t M{\"u}nchen},
    url = {https://mediatum.ub.tum.de/1741523},
}

@article{voglLiquidphaseLoopmodeArgon2024,
    title = {A Liquid-Phase Loop-Mode Argon Purification System},
    author = {Vogl, Christoph and Schwarz, Mario and Krause, Patrick and others},
    year = {2024},
    month = mar,
    journal = {J. Inst.},
    volume = {19},
    number = {03},
    pages = {C03030},
    issn = {1748-0221},
    doi = {10.1088/1748-0221/19/03/C03030},
}

@article{voglScintillationOpticalProperties2022,
    author = {Vogl, C. and Schwarz, M. and Stribl, X. and others},
    title = "{Scintillation and optical properties of xenon-doped liquid argon}",
    eprint = "2112.07427",
    archivePrefix = "arXiv",
    primaryClass = "physics.ins-det",
    doi = "10.1088/1748-0221/17/01/C01031",
    journal = "JINST",
    volume = "17",
    number = "01",
    pages = "C01031",
    year = "2022",
}

@mastersthesis{wiesingerTUMLiquidArgon2014,
    title = {The {{TUM}} Liquid Argon Test Stand: {{Commissioning}} and Characterization of a Low Background Test Stand for Background Suppression Studies in the Frame of the {{GERDA}} $0\nu\beta\beta$-Experiment},
    author = {Wiesinger, Christoph},
    year = {2014},
    address = {Garching},
    school = {Technische Universit{\"a}t M{\"u}nchen},
}

@article{darksidecollaborationResultsFirstUse2016,
    title = {Results from the First Use of Low Radioactivity Argon in a Dark Matter Search},
    author = {{DarkSide Collaboration} and Agnes, P. and others},
    year = {2016},
    month = apr,
    journal = {Physical Review D},
    volume = {93},
    number = {8},
    pages = {081101},
    publisher = {American Physical Society},
    doi = {10.1103/PhysRevD.93.081101},
}

@misc{GSTR-16-004,
    title = {Measurement of the {$^{42}$Ar} contamination in natural argon with {GERDA}},
    author = {K. von Sturm and R. Brugnera and S. Hemmer and others},
    year = {2016},
    note = {{GERDA} Collaboration, internal report}
}

@article{TARASOV20084657,
    title = {LISE++: Radioactive beam production with in-flight separators},
    journal = {Nucl. Inst. Meth. Phys. Res. B},
    volume = {266},
    number = {19},
    pages = {4657-4664},
    year = {2008},
    note = {Proceedings of the XVth International Conference on Electromagnetic Isotope Separators and Techniques Related to their Applications},
    issn = {0168-583X},
    doi = {10.1016/j.nimb.2008.05.110},
    author = {O.B. Tarasov and D. Bazin},
}

@article{PhysRevC.21.230,
    title = {Statistical model calculations in heavy ion reactions},
    author = {Gavron, A.},
    journal = {Phys. Rev. C},
    volume = {21},
    issue = {1},
    pages = {230--236},
    numpages = {0},
    year = {1980},
    month = {Jan},
    publisher = {American Physical Society},
    doi = {10.1103/PhysRevC.21.230},
}

@misc{PACE4,
    title = {{{PACE4}}},
    urldate = {2025-05-01},
    url = {https://lise.frib.msu.edu/pace4.html},
}

@article{Bethe,
    author = {Bethe, H.},
    title = {{Zur Theorie des Durchgangs schneller Korpuskularstrahlen durch Materie}},
    journal = {Annalen der Physik},
    volume = {397},
    number = {3},
    pages = {325-400},
    doi = {10.1002/andp.19303970303},
    year = {1930},
}

@article{Bloch,
    author = {Bloch, F.},
    title = {{Zur Bremsung rasch bewegter Teilchen beim Durchgang durch Materie}},
    journal = {Annalen der Physik},
    volume = {408},
    number = {3},
    pages = {285-320},
    doi = {10.1002/andp.19334080303},
    year = {1933},
}

\end{document}